\documentclass[amsmath,amssymb,showkeys,10pt]{revtex4}

\usepackage{hyperref} 

\usepackage{graphicx} 

\usepackage{bm}

\usepackage{amssymb}

\usepackage[a4paper]{geometry}

\newcommand{\Ai}{\mathrm{Ai}}
\newcommand{\kB}{k_\mathrm{B}}

\begin{document}

\title{Quantum Mechanical versus Stochastic Processes in Path Integration*}

\author{\href{https://www.etis.ee/Portal/Persons/Display/05ca5332-db93-42e6-a373-a15f0ebf1693?lang=ENG}{$~$Marco Patriarca$^{\,}_{\,}~$}\\
  {\it \href{www.kbfi.ee}{NICPB -- National Institute of Chemical Physics and Biophysics, R\"avala 10, 15042 Tallinn, Estonia}}\\
  {\small
    {\it E-mail}: {\tt marco.patriarca@kbfi.ee}
  } 
} 

\date{\today}


\begin{abstract} 
\noindent
{\bf *} Revised version of the contribution to \emph{PI'98}, the 6th International Conference on ``Path Integrals from peV to TeV: 50 years from Feynman's paper'', August 25th-29th 1998, Florence, Italy~\cite{Patriarca1999a}.
\vspace{0.3cm}\\
\noindent
{\bf Abstract}.
By using path integrals, the stochastic process associated to the time evolution of the quantum probability density $P(x,t) = |\psi(x,t)|^2$ is formally rewritten in terms of a stochastic differential equation, given by Newton's equation of motion with an additional multiplicative stochastic force.
However, the term playing the role of the stochastic force is defined by a non-positive-definite probability functional, providing a clear example of the negative (or ``extended'') probabilities characteristic of quantum mechanics.
%
\end{abstract}



\maketitle


\section{Introduction}
The present paper is motivated by the question, whether it is possible to recast quantum mechanics in the form of a stochastic process.
The question has been debated in the literature in various respects, such as the statistical interpretations of quantum mechanics, see e.g. Refs.~\cite{Grabert-1979a,Ballentine-1970a}.
In this paper it is shown that such a program cannot be accomplished in a straightforward way due to the appearance of ``negative probabilities'', i.e. of negative numbers where one would have expected positive probabilities, a known and intriguing feature of quantum mechanics, appearing whenever one tries to draw a direct analogy between quantum mechanics and statistical mechanics~\cite{Muckenheim-1986a,Muckenheim-1986b}.
Such non-positive ``probabilities'' can appear in various ways, typical examples being the non-positive extended probability distributions, such as the Wigner function, if interpreted as a probability density in the phase space of the particle~\cite{Muckenheim-1986b}.
Here,  non-relativistic quantum mechanics is formally rewritten in the form of a stochastic differential equation for the quantum particle coordinate.

\section{The effective Stochastic Process}
In 1982, Albert Schmid introduced a ``quasi-classical Langevin equation''~\cite{Schmid1982a} through a method that allows, as discussed below, to define a stochastic process associated to a quantum mechanical system.
Such a process will be referred to as the \emph{effective stochastic process} corresponding to a quantum mechanical one. 
The study in Ref.~\cite{Schmid1982a}, though concerned with quantum dissipative systems in a specific parameter range, contains ideas with a general validity, which are here applied to an isolated one-dimensional non-relativistic quantum system.
The quantum system can be described through the density matrix
$\rho(q, q', t) = \psi(q, t) \psi^*(q', t)$,
where $\psi(q, t)$ is the wave function of the system.
In the following, the focus will be on the time evolution of the corresponding probability density
$P(x, t) = \left. \rho(q, q', t) \right|_{q = q' = x}$.
The starting point is the introduction of the coordinates~\cite{Schmid1982a}
\begin{eqnarray}
\label{x-xi}
x = ( q + q' ) / 2 \, ,
~~~~
\xi = q' - q \, ,
\end{eqnarray}
and their interpretation as the observable value of position and the corresponding quantum fluctuation,
respectively --- see also Refs.~\cite{Schmid1982a,Illuminati1994a,Weiss-1999a}.
This can be proved by considering $\rho(q, q', t)$ as an infinite matrix,
with $q$ and $q'$ labeling rows and columns.
Then Eqs.~(\ref{x-xi}) define a new $x$-$\xi$ frame,
in which $\xi$ runs perpendicularly to the diagonal,
thus measuring quantum fluctuations~\cite{Weiss-1999a},
while $x$ runs along the diagonal and labels --- it is actually equal to --- the eigenvalues of position.

The density matrix of an isolated system, $\rho(q, q', t) = \psi(q, t) \psi^*(q', t)$, evolves with time according to
\begin{eqnarray}
\label{rho-1}
\rho(q_b, q_b', t_b) = 
\int dq_a \int dq_a' \, J(b|a) \rho(q_a, q_a', t_a) \, ,
\end{eqnarray}
through the propagator
\begin{eqnarray}
\label{J-1}
J(b|a) 
\equiv J(q_b, q_b', t_b | q_a, q_a', t_a)
=      K(q_b, t_b | q_a, t_a) K^*(q_b', t_b | q_a', t_a) \, ,
\end{eqnarray}
where the wave function propagator $K$ is given by [5]
\begin{eqnarray}
\label{K-1}
K(q_b, t_b | q_a, t_a) 
= \int_a^b Dq \, \exp\left( \frac{i}{\hbar} S[x] \right) \, .
\end{eqnarray}
Here $S[x]$ is the action of the system and
$a$ and $b$ are a short notation for the boundary conditions
$x(t_a) = x_a$ and $x(t_b) = x_b$, respectively.
For the action the following form is assumed,
\begin{eqnarray}
\label{S-1}
S[x] = 
\int_{t_a}^{t_b} dt \, 
\exp\left[ \frac{m}{2} \dot{q}^2(t) - V(q(t)) \right] \, ,
\end{eqnarray}
where $m$ is the particle mass and $V(q)$ the external potential.
Moving from the $q$-$q'$ to the $x$-$\xi$ frame through Eqs.~(\ref{x-xi}), one has
\begin{eqnarray}
\label{rho-2}
\rho(x_b, \xi_b, t_b) = 
\int dx_a \int d\xi_a \, \tilde{J}(b|a) \rho(x_a, \xi_a, t_a) \, ,
\end{eqnarray}
where the density matrix propagator $\tilde{J}(b|a)$ expressed in terms of the $x$ and $\xi$ variables is~\cite{Schmid1982a,Marinov-1980a,Marinov-1991a}
\begin{eqnarray}
\label{J-2}
\tilde{J}(b|a) 
\equiv \tilde{J}(x_b, \xi_b, t_b | x_a, \xi_a, t_a)
= \int Dx \int D\xi \, 
\exp\left\{
\frac{i}{\hbar} \int_{t_a}^{t_b} dt
\left[ 
- m \dot{x} \, \dot{\xi} 
+ \sum_{\pm} \pm V \left( x \pm \frac{\xi}{2} \right) 
\right]
\right\} .
\end{eqnarray}
Here and in the following the functional integrals over the variables $x$ and $\xi$ are understood to be made between the corresponding boundary conditions at times $t=t_a$ and $t=t_b$, i.e., $x(t_a) = x_a$, $\xi(t_a) = \xi_a$, and $x(t_b) = x_b$, $\xi(t_b) = \xi_b$. 
After integrating by parts the term proportional to $\dot{x} \, \dot{\xi}$
in Eq.~(\ref{J-2}) and letting $\xi_b = 0$, 
one obtains from Eq.~(\ref{rho-2}) the probability density $P(x_b, t_b) = \rho(x_b, \xi_b=0, t_b)$ at a generic time $t_b$ as
\begin{eqnarray}
\label{P-2}
P(x_b, t_b) = 
\int dx_a \, J_\mathrm{eff}(b|a) W_0(x_a,m\dot{x}_a,) \, ,
\end{eqnarray}
where the effective propagator $J_\mathrm{eff}(b|a)$ is given by
\begin{eqnarray}
\label{J-3}
J_\mathrm{eff}(b|a) 
= 
\int Dx \int D\xi \,
\exp\left\{
\frac{i}{\hbar} \! \int_{t_a}^{t_b} dt
\left[ 
m \xi \, \ddot{x} 
+ \sum_{\pm} \pm V \left( x \pm \frac{\xi}{2} \right) 
\right]
\right\} ,
\end{eqnarray}
and the function $W_0(x_a,m\dot{x}_a)$ is the Wigner function $W(x_a, p_a, t_a)$ at the initial time $t = t_a$ computed for $p_a = m \dot{x}_a$,
\begin{eqnarray}
\label{P-in}
W_0(x_a,m \dot{x}_a)
= 
\int d\xi_a \, \exp(i m \dot{x}_a \xi_a / \hbar) \, \rho(x_a, \xi_a, t_a) \, .
\end{eqnarray}
In general, $W_0(x_a,m \dot{x}_a)$ is not positive, thus introducing in turn already in the initial conditions a (well known) source of ``non-positive probability''.
However, in the following it is assumed that the initial Wigner function is positive (e.g. of Gaussian shape) to test whether, at least in those cases in which the initial conditions can be defined in terms of a positive probability density, it is possible to proceed toward a stochastic formulation of a quantum process.
It should also be remarked that while there is no well defined probability density of a quantum system in the configuration space $(\dot{x},x)$, here the probability density $P(x_b,t_b)$ in physical space in Eq.~\eqref{P-2} is considered, which is positive and well defined in quantum mechanics.
Equation \eqref{P-2}, together with Eqs.~\eqref{J-3} and~\eqref{P-in}, describe the effective stochastic process corresponding to the time evolution of the probability density $P(x,t) = |\psi(x,t)|^2$.

\section{Classical Limit}
\label{classical}

As a check on the formulas obtained above, it can be shown that they reproduce the classical limit of Newton's mechanics by an expansion in powers of $\xi$, representing the limit of small quantum fluctuations \cite{Illuminati1994a,Schmid1982a,Marinov-1980a,Marinov-1991a}.
To the second order in $\xi$, the effective propagator (\ref{J-3}) is
\begin{eqnarray}
\label{J-4}
J_\mathrm{eff}(b|a) 
\approx 
\int Dx \int  D\xi \,
\exp\left\{
\frac{i}{\hbar} \int_{t_a}^{t_b} dt \,
\xi \left[ 
m \ddot{x} + \partial_x V(x)
\right]
\right\} ,
\end{eqnarray}
where $\partial_x V(x) \equiv \frac{dV(x)}{dx}$.
In the polygonal approximation, the time interval $(t_a,t_b)$ is sliced into $N$ subintervals $(t_{k-1},t_k)$ of length $\varepsilon = (t_b - t_a)/N$ introducing the $N + 1$ discrete times $t_k = t_a + k \, \varepsilon$
($k = 0, \dots, N$; $t_0 \equiv t_a$; $t_N \equiv t_b$; $N \to \infty$)
and the functional integral is approximated by a product of integrals over the variables $x_k = x(t_k)$ and $\xi_k \equiv \xi(t_k)$,
$\int Dx \int D\xi \dots \to \prod_{k=1}^{N-1} \int dx_k \, d\xi_k \,\dots$
while $\int_{t_a}^{t_b} f(x(t)) \to (\varepsilon/2)\sum_{k = 1}^{N} [f(x_{k-1}) + f(x_{k})]$
where $f(x)$ is a generic function of the coordinate $x$.
Integrating over the variables $\xi_k$ in Eq.~(\ref{J-4}) gives the Dirac $\delta$-functions $\delta(m \ddot{x}_k + \partial_x V(x_k))$.
It is natural to generalize the $\delta$-function $\delta(x)$ through the ``$\Delta$-functional''
\begin{eqnarray}
\label{Delta}
  \Delta \left[ f \right]
  =
  \int D\xi \, \exp \left\{ 
                    i \int_{t_a}^{t_b} dt \, \xi(t) \, f(x(t)) 
                    \right\}
  \equiv 
  \lim_{N \to \infty} \prod_{k=1}^{N-1} \delta(f_k) \, .
\end{eqnarray}
Then Eq.~\eqref{J-4} becomes
\begin{eqnarray}
\label{J-5}
J_\mathrm{eff}(b|a) 
\approx 
\int_a^b Dx \,
\Delta \left[ m \ddot{x} + \partial_x V(x) \right] .
\end{eqnarray}
The $\Delta$-functional has a simple physical meaning~\cite{Crescimanno1993a}:
it selects the trajectory defined by setting its argument equal to zero,
since it assigns a zero probability to any other trajectories. 
Therefore, Eqs.~(\ref{J-5}) represents the classical propagator (in space $x$),
since it evolves the initial probability distribution along the classical trajectories $m \ddot{x} + \partial_x V(x) = 0$.

\section{Semi-classical Limit}
\label{semiclassical}

In this section we go one step further, by expanding the effective propagator,  Eq.~(\ref{J-3}), as far as the 4{\it th} order in $\xi$,
\begin{eqnarray}
\label{J-4b}
J_\mathrm{eff}(b|a) 
\approx 
\int Dx \int  D\xi \,
\exp\left\{
\frac{i}{\hbar} \int_{t_a}^{t_b} dt \,
  \left[ 
    \xi (m \ddot{x} + \partial_x V(x))
    + \frac{1}{24} \xi^3 \partial_x^3 V(x)
  \right]
\right\} .
\end{eqnarray}
In the polygonal approximation, one can rescale the variable $\xi_k$ in the generic integration term at time $t_k$, defining the variable
\begin{eqnarray}
  \label{eta}
  &&\eta_k = \xi_k \, \varphi(x_k) , \\
    \label{phi}
  &&\varphi(x) = \left( \frac{\partial_x^3 V(x)}{ 8 \hbar} \right)^{1/3} ,
\end{eqnarray}
so that Eq.~(\ref{J-4b}) becomes 
\begin{eqnarray}
\label{J-4c}
J_\mathrm{eff}(b|a) 
\approx
\lim_{N \to \infty}  \prod_{k=1}^{N-1} \int dx_k \int d\eta_k
\frac{1}{\varphi(x_k)}
\exp\left\{
  i \epsilon \,
  \left[ 
    \eta_k \left( \frac{m \ddot{x}_k + \partial_x V(x_k)}{ \hbar \varphi(x_k)} \right)
    +  \frac{1}{3} \eta_k^3 
  \right]
\right\} \, .
\end{eqnarray}
Because of the similarity between the integral in the $\eta$ variables and the Airy function~\cite{Abramowitz1970a},
\begin{equation}
  \Ai(x) = \int_{-\infty}^{+\infty} dy \exp[ i (xy + y^3/3) ] \, ,
\end{equation}
it is natural to introduce the ``Airy functional'' $\Ai[f]$ of a generic function $f(t)$ of time $t$,
\begin{equation}
  \label{AF}
  \Ai[f] = \int D\eta \exp \left\{ i \int_{t_a}^{t_b} \left[ f(t) \eta(t) + \frac{1}{3} \eta_k^3 \right] \right\}
           = \lim_{N \to \infty}  \prod_{k=1}^{N-1} \int d\eta_k \, \exp\left\{ i \epsilon \, \left[ \eta_k f(t_k) +  \frac{1}{3} \eta_k^3 \right] \right\}
  \, .
\end{equation}
Then Eq. (\ref{J-4c}), by defining the functional $\Phi[x] = \prod_k \varphi(x_k)$,  becomes
\begin{eqnarray}
\label{J-4d}
J_\mathrm{eff}(b|a) 
\approx 
\int_a^b Dx  \, \Phi[x]^{-1} \Ai [ (m \ddot{x} + \partial_x V) / \hbar \varphi(x) ] .
\end{eqnarray}
Finally, introducing the auxiliary process $R(t)$ and using the $\Delta$-functional given by Eq.~(\ref{Delta}), the effective propagator can be rewritten as
\begin{eqnarray}
\label{J-4e}
J_\mathrm{eff}(b|a) 
\approx 
\int DR \, \Ai[R] \int_a^b Dx  \,
  \Delta \left[ 
    m \ddot{x} + \partial_x V	(x) - \hbar \, \varphi(x) R
  \right] .
\end{eqnarray}
%

\section{Brownian Motion}
\label{stochastic}
%
The description obtained in the previous section for the (semi-classical) effective quantum propagator in physical space closely resembles the path-integral description of a classical Brownian particle~\cite{Risken-1989a}.
For the sake of simplicity, here the example of an overdamped Brownian particle subject to white noise in an environment at temperature $T$ is summarized.

A first way to describe such a Brownian particle is to use a Langevin equation, which defines the stochastic trajectories of the particle,
\begin{eqnarray}
\label{LE}
\frac{dx}{dt} =  - \frac{1}{m \gamma} \frac{\partial V(x)}{\partial x} + R(t) \, ,
\end{eqnarray}
where $\gamma$ is the friction coefficient and $R(t) = \xi(t)/m\gamma$ is a stochastic process, representing a white noise force $\xi(t)$ (divided by $m\gamma$), i.e., a Gaussian process with zero mean value and $\delta$-correlation,
\begin{eqnarray}
\label{R1}
\langle R(t) \rangle &=& 0 \, , \\
\label{R2}
\langle R(t) R(s) \rangle &=&  \frac{2 \kB T}{m \gamma} \delta(t - s)  \equiv 2 D \, \delta(t - s) \, ,
\end{eqnarray}
where $D$ is the diffusion coefficient.
Alternatively, one can describe the Brownian particle through its space-time dependent probability density $P(x,t)$.
The time evolution of $P(x,t)$ can be formulated as a drift-diffusion equation,
\begin{eqnarray}
\label{DE}
\frac{\partial P(x,t)}{\partial t} =  \frac{1}{m \gamma}  \frac{\partial}{\partial x} \left[ P(x,t) \frac{\partial V(x)}{\partial x} \right]  + D \frac{\partial^2 P(x,t)}{\partial x^2} \, .
\end{eqnarray}
The evolution law of $P(x,t)$ can be also be formulated in integral form,  
\begin{eqnarray}
  \label{PI}
  P(x_b,t_b) =  \int dx_a \, J(x_b, t_b|x_a,t_a) P(x_a,t_a)  \, ,
\end{eqnarray}
where the probability density propagator $J(b|a) = J(x_b, t_b|x_a,t_a)$ can be expressed as a path integral~\cite{Risken-1989a},
\begin{eqnarray}
\label{JWN}
J_\mathrm(b|a) 
= \int_a^b Dx \,  \exp\left\{
     - \frac{m\gamma\beta}{2} \int_{t_a}^{t_b} dt \, \left[
          \dot{x} + \frac{1}{m\gamma}\frac{\partial V(x)}{\partial x}
       \right]^2
   \right\} ,
\end{eqnarray}
The path integral formulation of Brownian motion is particularly interesting here, since it provides a link between (the time evolution of) the probability density and the Langevin equation (\ref{LE}) for the particle coordinate.
This is best seen by introducing the Gaussian functional
\begin{eqnarray}
\label{G}
G_D[R]  = \exp\left\{ - \frac{1}{2D}\ \int_{t_a}^{t_b} dt \, R(t)^2 \right\} ,
\end{eqnarray}
and using the $\Delta$ functional given above to rewrite the propagator (\ref{JWN}) as
\begin{eqnarray}
\label{JWN2}
J_\mathrm(b|a) = \int  dR \, G_D[R] \int_a^b Dx \, \Delta\left[ \dot{x} + \partial_x V(x)/m\gamma  -  R \right] \, .
\end{eqnarray}
This expression directly leads to the Langevin equation.
In fact, the functional integral over $x(t)$, through the $\Delta$-functional, selects the trajectory $\dot{x} = - \partial_x V(x)/m\gamma  +  R(t)$,
while the integral over $R(t)$ makes an average over different realizations of $R(t)$, weighting each trajectory with the functional $G_D[R]$, thus implicitly defining $R(t)$ as a Gaussian stochastic process with first  moment $\langle R(t) \rangle = \int dR \, R(t) \, G_D[R] = 0$ and second moment $\langle R(t) R(s) \rangle = \int dR \, R(t) \, R(s)  \, G_D[R] = 2 D \delta(t - s)$. This provides a description equivalent to that of Eqs.~(\ref{LE})-(\ref{R2}).

\section{Comparison between quantum and stochastic process}
\label{comparison}
\begin{figure}[!ht]
\begin{center}
 \includegraphics[width=10.0cm]{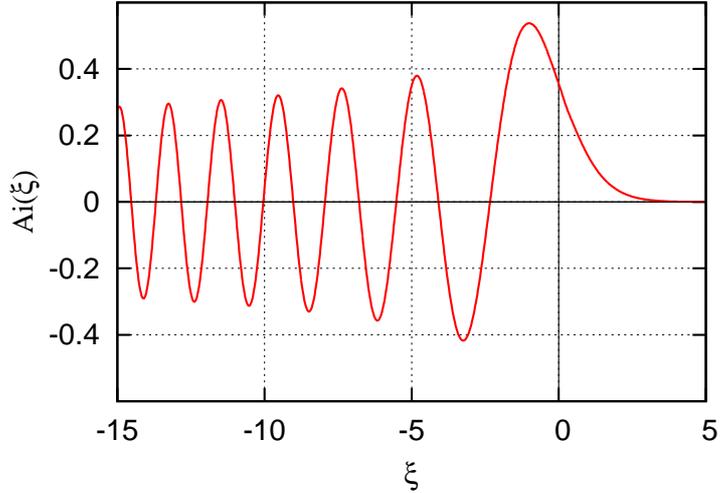}
\end{center} 
\caption{
Plot of the Airy function $\Ai(\xi)$~\cite{Abramowitz1970a} showing some interval of the $\xi$-axis where $\Ai(\xi) < 0$.
\label{fig_airy}
 }
\end{figure}

%
One can now compare Eqs.~(\ref{JWN2}) and (\ref{J-4e}), to conclude that the quantum propagator is given by a sum of contributions from particle trajectories described by the equation 
\begin{eqnarray}
\label{QLE}
m \ddot{x} = - \partial_x V	(x) + \hbar \, \varphi(x) R \, ,
\end{eqnarray}
which is Newton's equation of motion with the additional force term  $\hbar \, \varphi(x(t)) R(t)$, where  $\varphi(x)$ is given in Eq.~(\ref{phi}), with each trajectory weighted by the Airy functional $\Ai[R]$ in Eq.~(\ref{AF}).
However, the additional force term cannot be considered as a multiplicative stochastic force, since, on the contrary of the case of a Brownian particle, here the Airy function $\Ai(\xi)$ (and therefore the Airy functional $\Ai[f]$) is not positive-definite~\cite{Abramowitz1970a}, as it would be expected for a probability density function(al), and assumes negative values in some ranges of the semi-axis $\xi < 0$, see Fig.~\ref{fig_airy}.

The non-positivity of the Airy function(al) prevents an operative definition of $R(t)$ as a stochastic process as well as the possibility of performing a numerical simulation of the trajectories defined by Eq.~(\ref{QLE}).

\section{Conclusion}
\label{conclusion}
%
The path-integral formalism of quantum mechanics is a natural framework to connect the differential (Langevin) formulation of Brownian motion and its corresponding (path) integral formulation.
When trying to use path integrals to reformulate in an analogous way (the semi-classical limit of) quantum mechanics in the form of a stochastic differential equation for the particle coordinate, however,  one finds a differential equation containing a term playing the role of a ``stochastic force'' associated to a non-positive-definite probability distribution function.
This result provides a clear example of the negative probabilities appearing in quantum mechanics~\cite{Muckenheim-1986b} and confirms previous findings suggesting that any statistical interpretation of quantum mechanics cannot be based on a plain equivalence between quantum mechanical and classical stochastic processes.


\begin{center}
{\bf ---------------------------------------------------------------}
\end{center}

\begin{acknowledgments}
This revision of Ref.~\cite{Patriarca1999a} was made possible by the support of the European Regional Development Fund (ERDF) Center of Excellence (CoE) program grant TK133 and the Estonian Research Council through Institutional Research Funding Grants (IUT) No. IUT-39-1, IUT23-6, and Personal Research Funding Grant (PUT) No. PUT-1356.
\end{acknowledgments}


\newpage


\begin{thebibliography}{10}
\newcommand{\enquote}[1]{``#1''}

\bibitem{Patriarca1999a}
M.~Patriarca.
\newblock \enquote{Quantum mechanical versus stochastic processes in path
  integration.}
\newblock In R.~Casalbuoni, R.~Giachetti, V.~Tognetti, R.~Vaia, P.~Verrucchi
  (editors) \enquote{Path Integrals from peV to TeV: 50 years from Feynman's
  Paper,} pp. 589--592. World Scientific, Singapore, 1999.

\bibitem{Grabert-1979a}
H.~Grabert, P.~H\"anggi, P.~Talkner.
\newblock \enquote{Is quantum mechanics equivalent to a classical stochastic
  process?}
\newblock \emph{Phys. Rev. A}, vol.~19, pp. 2440--2445, 1979.

\bibitem{Ballentine-1970a}
L.~E. Ballentine.
\newblock \enquote{The statistical interpretation of quantum mechanics.}
\newblock \emph{Rev. Mod. Phys.}, vol.~42, p. 358, 1970.

\bibitem{Muckenheim-1986a}
W.~M{\"u}ckenheim.
\newblock \enquote{What is an extended probability?}
\newblock \emph{Nature}, vol. 324, p. 307–307, 1986.

\bibitem{Muckenheim-1986b}
W.~M{\"u}ckenheim.
\newblock \enquote{A review of extended probabilities.}
\newblock \emph{Phys. Rep.}, vol. 133, pp. 337--401, 1986.

\bibitem{Schmid1982a}
A.~Schmid.
\newblock \enquote{On a quasi-classical {L}angevin equation.}
\newblock \emph{J. Low Temp. Phys.}, vol.~49, p. 609, 1982.

\bibitem{Illuminati1994a}
F.~Illuminati, M.~Patriarca, P.~Sodano.
\newblock \enquote{Classical and quantum dissipation in nonhomogeneous
  environments.}
\newblock \emph{Physica A}, vol. 211, p. 449, 1994.

\bibitem{Weiss-1999a}
U.~Weiss.
\newblock \enquote{Dissipative quantum systems.}
\newblock In R.~Casalbuoni, R.~Giachetti, V.~Tognetti, R.~Vaia, P.~Verrucchi
  (editors) \enquote{Path Integrals from peV to TeV: 50 years from Feynman's
  Paper,} World Scientific, Singapore, 1999.

\bibitem{Marinov-1980a}
M.~Marinov.
\newblock \enquote{Path integrals in quantum theory: An outlook of basic
  concepts.}
\newblock \emph{Physics Reports}, vol.~60, pp. 1 -- 57, 1980.

\bibitem{Marinov-1991a}
M.~Marinov.
\newblock \enquote{A new type of phase-space path integral.}
\newblock \emph{Physics Letters A}, vol. 153, pp. 5 -- 11, 1991.

\bibitem{Crescimanno1993a}
M.~Crescimanno.
\newblock \enquote{Quantum-mechanics and thermal noise in dissipative systems.}
\newblock \emph{Annals of Physics}, vol. 223, pp. 229 -- 242, 1993.

\bibitem{Abramowitz1970a}
M.~Abramowitz, I.~A. Stegun (editors) \emph{Handbook of Mathematical
  Functions}.
\newblock Dover, N.Y., 1970.

\bibitem{Risken-1989a}
K.~Risken.
\newblock \emph{The Fokker-Planck Equation}, vol. Chap. 4.4.2, p. 74.
\newblock Springer-Verlag, Berlin, 1989.

\end{thebibliography}


\end{document}